\documentstyle[prb,eqsecnum,preprint,aps]{revtex}

\begin{document}

\title{Nonlocal Conductivity in Type-II Superconductors}

\author{Chung-Yu Mou$^1$, Rachel Wortis$^{2,3}$, Alan T. Dorsey$^1$ and David
A. Huse$^3$}

\address{
$^1$Department of Physics, University of Virginia,  \\
 McCormick Road, Charlottesville, Virginia 22901}

\address{
$^2$ Department of Physics,
 University of Illinois at Urbana-Champaign, \\
1110 West Green Street, Urbana, Illinois 61801 }

\address{
$^3$AT\&T Bell Laboratories, Murray Hill, NJ 07974 }

\date{\today}

\maketitle

\begin{abstract}

Multiterminal transport measurements on YBCO crystals in the vortex
liquid regime have shown nonlocal conductivity on length scales
up to 50 microns.  Motivated by these results we explore the
wavevector ({\bf k}) dependence of the dc conductivity tensor,
$\sigma_{\mu\nu} ({\bf k})$, in the Meissner, vortex lattice, and
disordered phases of a type-II superconductor.
Our results are based on time-dependent Ginzburg-Landau (TDGL) theory
and on phenomenological arguments.  We find four qualitatively different
types of behavior.  First, in the Meissner phase, the conductivity is
infinite at $k=0$ and is a continuous function of $k$,
monotonically decreasing with increasing $k$.  Second, in the
vortex lattice phase, in the absence of pinning, the conductivity is
finite (due to flux flow) at $k=0$;  it is discontinuous there
and remains qualitatively like the Meissner phase for $k>0$.  Third,
in the vortex liquid regime in a magnetic field and at low
temperature, the conductivity is finite, smooth and {\it non-monotonic},
first increasing with $k$ at small $k$ and then decreasing at larger $k$.
This third behavior is expected to apply at temperatures just above the
melting transition of the vortex lattice, where the vortex liquid shows
strong short-range order and a large viscosity.
Finally, at higher temperatures in the disordered phase, the conductivity is
finite,
smooth and again monotonically decreasing with $k$.  This last, monotonic
behavior applies
in zero magnetic field for the entire disordered phase, i.e. at all
temperatures
above $T_c$, while in a field the non-monotonic behavior may occur in a low
temperature portion of the disordered phase.

\end{abstract}


\section{Introduction}

In this paper we explore the nonlocal dc electrical
transport properties of type-II superconductors.
What is meant by nonlocal here?  The general expression connecting the
local current density in a material, ${\bf J}$, to the local electric field,
${\bf E}$,
in the linear (Ohmic) regime is
\begin{equation}
J_{\mu}({\bf r})=\int \sigma_{\mu \nu}({\bf r},{\bf r'})
E_{\nu}({\bf r'})\,d{\bf r'} .
\label{nonlocal1}
\end{equation}
If the conductivity ${\bf \sigma(r,r')}$ is nonvanishing for ${\bf r \neq r'}$,
then this is nonlocal;  the current at ${\bf r}$ is determined by the field
not only at ${\bf r}$, but at all points ${\bf r'}$ where ${\bf \sigma(r,r')}
\neq 0$.
In a translationally invariant system, the nonlocal conductivity can only be a
function of the difference $({\bf r}-{\bf r'})$.  Taking the Fourier
transform one then obtains
$J_{\mu}({\bf k})=\sigma_{\mu\nu}({\bf k})E_{\nu}({\bf k})$.
All materials exhibit nonlocal transport properties on some length scale.
In normal metals,
the nonlocal terms are significant only at length scales less than or
of the order of the inelastic mean free path.  In superconducting materials,
however, as the transition is approached from above,
the associated correlations can cause
nonlocal effects to become important over much longer length scales.
It is this phenomenon that we address in this paper.

Vortices are important actors in the nonlocal electrical transport properties
in type-II superconductors.  This is sometimes more easily described
in terms of the nonlocal dc resistivity: \cite{huse93}
\begin{equation}
E_{\mu}({\bf r})=\int \rho_{\mu \nu}({\bf r,r'})J_{\nu}({\bf r'})\,d{\bf r'}  .
\label{nonlocres}
\end{equation}
(Note that the nonlocal resistivity is a linear operator that acts on the full
current pattern, ${\bf J(r)}$, and is, as usual, the inverse of the
nonlocal conductivity operator.)
A current ${\bf J(r')}$ pushes on the vortex segments at ${\bf r'}$
due to the Lorentz and Magnus forces.  These vortex segments  move,
and, due to the continuity and entanglement of vortices as well as the
repulsive and attractive forces between parallel and antiparallel
(respectively)
vortices, they cause vortex segments at ${\bf r}$ to move.\cite{marchetti90}
This
produces phase slip and electric fields at ${\bf r}$.  Thus vortex
dynamics contribute significantly to the nonlocal resistivity,
$\rho_{\mu \nu}({\bf r,r'})$.

Recent multiterminal transport experiments \cite{safar92} on YBCO
crystals in the vortex liquid regime obtain results that demonstrate
nonlocal conductivity on length scales of at least 50 microns.
A phenomenological understanding of those experimental results was
presented\cite{huse93}
based on a hydrodynamic description of the vortex liquid and its
viscosity.  In this paper we also examine the
vortex lattice phase and extend the phenomenological study of the vortex liquid
regime to low and zero magnetic field.  We then
calculate the nonlocal conductivity due to the Gaussian
fluctuations above $T_c$ within time-dependent Ginzburg-Landau (TDGL) theory.
An outline of the paper is as follows:

Section II considers an ideal, unpinned, defect-free vortex lattice.
There the steady-state motion of the vortices is only a rigid-body (flux-flow)
motion of the vortex lattice as a whole.  The electric field due to this vortex
motion
is therefore determined only by the total
force and torque on the entire vortex lattice.  A dc current pattern with
wavevector $k \neq 0$ elastically distorts the lattice, but, in linear
response,
produces no steady motion of the vortices. Thus we find that in this ideal case
the Ohmic flux-flow resistivity due to vortex motion is nonzero only at $k=0$.
In both the vortex lattice and Meissner phases there is an additional
contribution to the resistivity that is proportional to $k^2$ at small $k$
and does not arise from vortex motion.\cite{schmidt70}

In Section III we discuss the phenomenology of the vortex liquid regime.
Connectivity, entanglement and other interactions between nearly parallel
vortices
then give rise to the vortex-liquid viscosity\cite{marchetti90} that
reduces their motion in
response to nonuniform dc currents, producing a resistivity that
has a local maximum for uniform ($k=0$) currents and is smaller for
long-wavelength nonuniform ($k \neq 0$) currents, as in the
case of the vortex lattice.
However, at high enough temperature and low or zero magnetic field
there will be thermally excited vortex lines and loops present with
all orientations.  Then the connectivity and attraction of nearly antiparallel
vortices
produces a nonlocal effect of the opposite sign, with the resistivity
smallest at $k=0$.  In the disordered phase, the conductivity and the
resistivity may be expanded in powers of $k$;
these arguments then indicate that the signs
of the order $k^2$ terms will depend on temperature and magnetic field.

Section IV sets up the TDGL calculation of the lowest-order (Gaussian
approximation) fluctuation contribution to the nonlocal dc conductivity
in the disordered phase.  For a uniform ($k=0$) current this Gaussian
approximation gives the Aslamazov-Larkin fluctuation
conductivity.\cite{aslamazov68}
The calculation for zero magnetic field is carried out in Section V.
Here the full $k$-dependence can be obtained and the conductivity is
found to be a monotonically decreasing function of $k$.  Section VI
obtains the fluctuation conductivity to order $k^2$ in a magnetic
field.  At this order, the sign of the $k$-dependence remains the
same as for zero magnetic field, at least for longitudinal electric fields.
We conclude with a brief summary and discussion in Section VII.
Appendices A and B are devoted to some technical details of the TDGL
calculation.

\section{Vortex Lattice}

Let us first consider the vortex lattice phase.  In dc steady state the
total time-averaged force on each portion of the vortex
lattice, including drag forces, must
vanish, since the lattice is either stationary or moving with a finite
steady-state velocity.  In linear response to a dc current, the
resulting steady-state local force
balance equation for the vortex lattice contains four terms:  one
proportional to and perpendicular to the local current
density ${\bf J}$ (arising from Lorentz and/or Magnus forces),\cite{vinen66}
one proportional to and perpendicular to the local vortex
velocity ${\bf v}$ (arising from the Magnus force and/or a perpendicular drag
force), one proportional to and parallel to the local vortex velocity (a
standard
drag force), and one arising from the local elastic distortion of the
vortex lattice.\cite{larkin86}  Assuming the magnetic induction
${\bf B} = B \hat{\bf z}$, so that the vortices are
parallel to the $z$-axis,  and that the material is isotropic in
the $xy$-plane, the steady state force balance equation is
\begin{equation}
{\bf J \times B} + \gamma_2 n {\bf \hat z \times v} - \gamma_1 n {\bf v}
-{\delta H_{elastic} \over
\delta {\bf u}} = 0 ,
\label{lattice1}
\end{equation}
where
\begin{equation}
H_{elastic}={1 \over 2} \sum_{\bf k} u_i(-{\bf k})
\bigl\{ c_L({\bf k}) k_i k_j + \delta_{ij}[c_{66}({\bf k}) k_{\perp}^2
+ c_{44}({\bf k}) k_{z}^2]\bigr\} u_j({\bf k})
\label{lattice2}
\end{equation}
is the elastic energy of the vortex lattice.\cite{houghton89}
${\bf u}$ is the local displacement of the vortices away from an ideal,
undistorted lattice, and  $c_L$, $c_{66}$ and $c_{44}$ are
the bulk, shear and tilt elastic moduli, respectively.
Here ${\bf v}$ refers to the local velocity of the vortices
in the steady state, averaged over time and over a length scale longer
than the lattice spacing.  We assume the lattice is dislocation free
so that ${\bf u}$ is well defined, and that there are no vacancies or
interstitials so that ${\bf v}$ is simply $d{\bf u}/dt$.
The areal density $n$ is related to the
magnetic induction by $n=B/\phi_{0}$, with $\phi_{0}=h/2e$ the flux
quantum.  The drag coefficients $\gamma_{1}$ and $\gamma_{2}$
are phenomenological parameters.

When the current is spatially uniform, the resulting vortex velocities and
displacements are also spatially uniform.  The elastic term in
(\ref{lattice1}) is then zero and we are left with a balance of the
other three terms.
Using the Josephson relation for the electric field produced by
moving vortices,\cite{josephson66}
\begin{equation}
{\bf E} = - {\bf v \times B},
\label{lattice3}
\end{equation}
one finds for the components of the conductivity tensor
\begin{eqnarray}
\sigma_{xx}&=& \sigma_{yy}= {\gamma_1 \over \phi_{0} B}\\
\label{lattice6}
\sigma_{xy}&=& -\sigma_{yx}= {\gamma_2 \over \phi_{0} B}  ,
\nonumber
\end{eqnarray}
$\sigma_{zz}$ is infinite and the other components are zero.
The effect is that when a spatially uniform current
is applied, the vortex lattice as a whole is
pushed across the sample, causing phase
slip and therefore dissipation.  This is the standard
flux-flow resistivity.

Now suppose the current is nonuniform.  The elastic term is then nonzero
because different parts of the vortex lattice are subject to
different forces and this results in elastic strains.
If the current produces a nonzero total force or torque on the vortex
lattice, it will move as a (overdamped) rigid body in response, again
exhibiting flux-flow resistivity.
However, if the spatially averaged current
(as well as the total torque applied to the vortex lattice by the current)
vanishes, there is no steady-state motion of the vortex lattice.  Instead
the lattice is statically strained so that
everywhere the elastic restoring force
locally balances the force due to the nonuniform current.
The vortices are not moving, so no electric field is being produced,
although a current is flowing.
Therefore, the resistivity due to vortex motion vanishes in linear response to
a
long-wavelength,
static, nonuniform current pattern with zero spatial average.
For example, if a dc current is applied in the $x$ direction with variation in
the $x$ direction, then the vortices will be pushed along the $y$ direction,
producing static shear elastic distortions.  Such an example is illustrated
in Fig. 1.

Here the dc resistivity due to vortex motion is completely nonlocal, since
there
is only rigid body motion of the vortex
lattice as a whole, which is determined by the total force (and
torque) applied to the entire vortex lattice by the current.  Of course,
this is a very idealized discussion, since we are neglecting pinning,
defects in the vortex lattice (and the resulting plastic motion),
and the boundaries of the sample.  However, this shows that in this
ideal case, the freezing of the vortices into a lattice is indeed a
superconducting transition.  Although the flux-flow resistivity to a
uniform ($k=0$) current does not vanish when the vortices freeze, the dc
flux-flow resistivity to a long-wavelength nonuniform ($k > 0$) current does
vanish.

We have examined the response of the vortex lattice to a nonuniform
electric field within the time-dependent Ginzburg-Landau (TDGL) equations
in the absence of thermal noise, following, e.g., Troy and
Dorsey.\cite{troy93}
The above phenomenology for the resistivity due to vortex motion is confirmed,
but for nonuniform currents there is an additional
contribution to the resistivity which does
not arise from vortex motion.  This dissipation in a static, but strained
configuration arises directly from the gradient-squared term in the TDGL
equations
and gives a resistivity proportional to $k^2$ for small wavevector $k$
in both the vortex lattice and Meissner phases.\cite{schmidt70}

To summarize, the resistivity in the ideal vortex lattice phase is nonzero
at $k=0$ and proportional to $k^2$ for small nonzero $k$.  It is therefore
discontinuous and non-monotonic in $k$ at $k=0$.  We expect that this
discontinuity and non-monotonicity is still present if the lattice contains
a non-zero density of vacancies and interstitials: then the resistivity for
$k=0$ is due to motion of the entire lattice, while for small nonzero $k$
only the defects move, resulting in a lower resistivity.

\section{Vortex Liquid Phenomenology}

Let us now consider the vortex liquid regime.
This is the regime in which Huse and Majumdar\cite{huse93} constructed their
phenomenological theory for the conductivity measurements of Safar {\it et
al.}.\cite{safar92}  They begin with a force balance equation including
three terms:  one proportional to the current density, J (arising
from Lorentz and/or Magnus forces), one proportional to the average
vortex velocity and hence the electric field (arising from drag and Magnus
forces), and one proportional to the second spatial derivative of the
vortex velocity and therefore the second spatial derivative of the
electric field (arising from viscous forces):
\begin{equation}
J_{\mu}({\bf r}) = \sigma_{\mu \nu}(0) E_{\nu}({\bf r}) -
S_{\mu \alpha \beta \nu} \partial_{\alpha} \partial_{\beta}
E_{\nu}({\bf r}).
\label{liquid1}
\end{equation}
Although this equation was initially motivated by considering only
the electric field due to vortex motion, it is more generally just
the long-wavelength expansion of the nonlocal Ohm's law (1.1).
The nonlocal conductivity to order $k^2$ is then
\begin{equation}
\sigma_{\mu \nu}({\bf k})=\sigma_{\mu \nu}(0) +
S_{\mu \alpha \beta \nu} k_{\alpha} k_{\beta}.
\label{liquid2}
\end{equation}
Note that in this paper we call the coefficient of the last
term $S$, not $\eta$ as in
Huse and Majumdar,\cite{huse93} in order to avoid confusion with
either the Bardeen-Stephen drag coefficient or the vortex liquid
viscosity tensor.

How is $S$ related to the
hydrodynamic viscosity tensor of the vortex liquid?
This can be answered for the components of $S$ that couple to
electric fields in the $xy$ plane using the
more detailed force balance equation,
similar to that discussed above for the
lattice, given in Marchetti and Nelson.\cite{marchetti90}
Again, ${\bf B}=B \bf \hat z$.  Neglecting compressibility as well
as vortex segments
which are not parallel to the $z$-axis and any hexatic
bond-orientational order, the only difference between the
dc force balance equation for the liquid and the
lattice is that the elastic force is replaced by a viscous force:
\begin{equation}
B({\bf J \times \hat z})_i + \gamma_2 n ({\bf \hat z \times v})_i
- \gamma_1 n v_i +{1 \over 2}\eta_{ijkl}\bigl( \partial_j \partial_k v_l
+ \partial_j \partial_l v_k \bigr) = 0  .
\label{liquid3}
\end{equation}
A note on notation:  Bardeen and Stephen refer to the coefficient of the
drag term as $\eta$.  Here and in Marchetti and Nelson,\cite{marchetti90}
the drag coefficients are $\gamma$'s, and $\eta$ is the viscosity,
which enters as the coefficient of the
$\nabla^2$ term.  Note also that this $\eta$ involves
interactions between vortices, as well as the connectivity
and entanglement of vortex lines.  If we neglect any dissipation
that is not associated with vortex motion,
following the same steps as for the lattice, we obtain
\begin{equation}
\sigma_{xx}({\bf k}) = {1 \over \phi_{0} B} \bigl[ \gamma_1 +
\phi_{0} \eta_{y \alpha y \beta} k_{\alpha}k_{\beta}
\bigr], \hspace{0.1in}
\sigma_{xy}({\bf k}) = {1 \over \phi_{0} B} \bigl[ \gamma_2 -
\phi_{0} \eta_{y \alpha x \beta} k_{\alpha}k_{\beta} \bigr],
\mbox{\hspace{0.1in} \rm etc.}
\label{liquid4}
\end{equation}
Expressions for the resulting hydrodynamic contributions to $S$
can be read off from these equations.

However, the approximation of neglecting vortex segments not running
parallel to the $z$-axis is inappropriate in a vortex liquid whose
uniform ($k=0$)
resistivity parallel to the $z$-axis is nonzero.  This resistivity
is due to the motion of precisely those vortex segments that do
not run parallel to the $z$-axis.  Thus more generally we should consider
a vortex liquid containing vortices running in any direction.  In a
large enough magnetic field and at low enough temperatures, only the
field-induced vortices are present, and they are all nearly parallel
to each other and to the $z$-axis.  When a uniform current is applied
perpendicular to the vortices they all move in the same direction, as
in the case of the vortex lattice.  However in a nonuniform ($k=0$) current
vortices in neighboring regions experience different forces.  Unlike
in the vortex lattice the resulting shears are not simply balanced by
elastic forces, instead the vortices do continue to move past each other.
However, their motion is slowed (relative to the case of a uniform current)
by the vortex-liquid viscosity arising from connectivity,
entanglement and other vortex-vortex interactions.
Slower vortex motion means reduced flux-flow resistivity.  Thus
the flux-flow resisitivity decreases with increasing $k$ and the conductivity
increases.  This corresponds to positive viscosities in the above
hydrodynamic model and gives, for example, $S_{xxxx} > 0$.

At higher temperatures and in low or zero magnetic field, there will
be thermally excited vortex line segments present with all orientations
that outnumber the field-induced vortices.  Parallel vortices interact
repulsively, while antiparallel vortices attract each other.  Thus the
near neighbors to a given vortex segment are more likely to be antiparallel.
Such nearby antiparallel segments can be joined into a vortex loop, or,
in a film can form a bound vortex-antivortex pair.  When a
current is applied, those segments perpendicular to the
current will feel the resulting Magnus/Lorentz force.
If the current is in the $x$ direction, a
vortex segment parallel to positive ${\bf \hat z}$ will be pushed in the
negative y direction while a segment parallel to negative ${\bf \hat z}$
will be pushed in the positive $y$ direction.  Both motions induce
electric fields of the same sign along the $x$ direction and thus contribute to
the
flux-flow resistivity.  However, the relative motion of such antiparallel
vortices is impeded by their being connected
or entangled, {\it as well as} by the attractive force between
antiparallel vortices.  In a uniform ($k=0$) current the antiparallel
segments feel equal and opposite forces from the current,
so this is when their motion
is most impeded by connectivity, entanglement or other interactions.
For a nonuniform ($k>0$) current, the forces do not cancel and part of the
motion generated is ``center-of-mass'' instead of relative, so is not
impeded by the interactions.  Thus we expect more vortex motion for $k>0$,
in this case where it is the relative motion of antiparallel vortices
that dominates the resistivity.  More vortex motion means a larger
electric field, greater resistivity and
lower conductivity.  Thus here the conductivity for small k is maximal for
uniform ($k=0$) current and decreases with increasing k.  This
corresponds to negative viscosities in the above hydrodynamic model
and gives for example $S_{xxxx} < 0$.  Thus we expect that
the sign of the nonlocal effect (the $S_{ijkl}$'s) will change
as one varies the field and/or temperature in the vortex liquid regime.
This sign change has been confirmed for $S_{xxxx}$ in preliminary Monte
Carlo simulations of a simple two-dimensional model
superconductor.\cite{huse94}

The above phenomenological arguments deal with the long-wavelength (small $k$)
behavior.  At short wavelengths (large $k$) we expect the qualitative behavior
is
not sensitive to the various phase transitions and distinctions that
affect the long-wavelength behavior.
In all the regimes where we can
obtain the large-$k$ behavior, namely, the Meissner and vortex lattice
phases in TDGL without fluctuations and the treatment below of the fluctuation
conductivity in TDGL in the zero-field normal state, we find the conductivity
decreases
with increasing $k$ in the large-$k$ regime.  Thus we expect this remains
true throughout the vortex liquid as well.  This then suggests that when the
$k$-dependence at small $k$ changes sign, the conductivity is changing from
monotonic in $k$ (in the higher-temperature, lower-field regime),
to non-monotonic in $k$ (in the lower-temperature, higher-field regime).

\section{The time-dependent Ginzburg-Landau equations and linear response}

In order to calculate the nonlocal conductivity of a superconductor
we need to specify the dynamical equations of motion for the
superconducting order parameter $\psi$.  We will adopt the simplest such
description, the time-dependent Ginzburg-Landau (TDGL) equation:
\begin{equation}
\Gamma^{-1} (\partial_{t} + i {e^{*}\over \hbar} \phi) \psi =
{\hbar^{2} \over 2 m} (\nabla - i {e^{*}\over \hbar} {\bf A} )^{2} \psi
- a \psi - b|\psi|^{2} \psi + \zeta,
\label{tdgl1}
\end{equation}
where $\phi$ is the scalar potential, $m$ is the effective mass of a
Cooper pair, $e^{*}=2e$ is the
charge of a Cooper pair, $a(T)=a_{0}(T/T_{c}-1)$ with $T$ the temperature and
$T_{c}$ the zero-field
critical temperature in the absence of the noise term, and $\Gamma$ is the
order parameter
 relaxation rate (taken to
be real).  Note that we are treating an isotropic superconductor.
The stochastic noise term $\zeta({\bf x},t)$ is chosen to have
Gaussian white noise correlations, with the two-point correlation function
given by
\begin{equation}
\langle \zeta^{*} ({\bf x},t)\zeta ({\bf x'},t')\rangle = 2 \Gamma^{-1}
k_{B} T \delta^{(d)}({\bf x} - {\bf x}') \delta(t-t'),
\label{noise1}
\end{equation}
with the coefficient being determined by the fluctuation-dissipation
theorem.\cite{hohenberg77}  We will work in the limit of large Ginzburg-Landau
parameter, $\kappa=\lambda/\xi$, where we can neglect the fluctuations in
the magnetic field.  Thus the vector potential, ${\bf A}$, is static and
is simply that due to a uniform magnetic field, and ${\bf H}={\bf B}$.

As we are interested in the linear response of the system to an applied
electric field, we can calculate
the conductivity matrix by using the
Kubo formula, which expresses the conductivity as the Fourier transform
of the current-current correlation function:
\begin{equation}
\sigma_{\mu\nu}({\bf k},\omega) = {1\over 2 k_{B}T}\int d^{d}({\bf x}-{\bf x}')
\int d(t-t') e^{i{\bf k}\cdot({\bf x}-{\bf x}') -i\omega (t-t')}
\langle J_{\mu}({\bf x},t) J_{\nu}({\bf x}',t')\rangle.
\label{kubo}
\end{equation}
Since we are ignoring fluctuations in the vector potential, the current
which appears in the Kubo formula is the supercurrent,
\begin{equation}
{\bf J_s} = {\hbar e^* \over 2 m i}(\psi^*\nabla\psi - \psi\nabla\psi^*)
 - {(e^*)^2 \over m}|\psi|^2{\bf A},
\label{js}
\end{equation}
and the conductivity
is due to superconducting fluctuations; the total conductivity is obtained
by adding this contribution to the normal state conductivity.
The validity of the Kubo formula in the context of the TDGL equations (with
real $\Gamma$)
 can be demonstrated.\cite{mou94}  However, when $\Gamma$ is complex,
the usual form of the Kubo formula may need
to be modified.\cite{mou94}

If we make the {\it Gaussian approximation} and neglect the cubic
term in the TDGL equation, Eq.\ (\ref{tdgl1}), then the
current-current correlation function factors into a product of
order parameter correlation functions, with the result that
\begin{eqnarray}
\sigma_{\mu\nu}({\bf k})&=&{1\over 2 k_{B} T} \left({\hbar e^{*}\over 2 m i}
 \right)^{2} \int {d\omega\over 2\pi} \int d^{d}({\bf x}_{1}-{\bf x}_{2})
 e^{i{\bf k}\cdot({\bf x}_{1}-{\bf x}_{2})} \nonumber \\
& & \qquad (D_{\mu 1} - D_{\mu 3}^*) (D_{\nu 2} - D_{\nu 4}^*) \left.
 C_{0}({\bf x}_{2},{\bf x}_{3}, \omega)  C_{0}({\bf x}_{1},{\bf x}_{4}, \omega)
 \right|_{3=1,4=2},
\label{gaussian}
\end{eqnarray}
where $D_{\mu}=\partial_{\mu}-ie^{*}A_{\mu}/\hbar$, and
\begin{eqnarray}
C_{0}({\bf x},{\bf x}';\omega) &=& \int {d\omega \over 2\pi}e^{-i\omega(t-t')}
    \langle \psi({\bf x},t) \psi^{*}({\bf x}',t') \rangle \nonumber \\
    &=& {2 k_{B} T\over \omega} {\rm Im} G_{0}({\bf x},{\bf x}';\omega),
\label{correlation1}
\end{eqnarray}
with $G_{0}({\bf x},{\bf x}';\omega)$ the order parameter response function,
which is the solution to
\begin{equation}
\left[ -i\Gamma^{-1} \omega -{\hbar^{2} \over 2 m} (\nabla
   - i {e^{*}\over \hbar} {\bf A} )^{2} + a\right]
  G_{0}({\bf x},{\bf x}';\omega)= \delta^{(d)} ({\bf x}-{\bf x}').
\label{correlation2}
\end{equation}
Within the Gaussian approximation we have also obtained the conductivity
directly from the equation of motion (\ref{tdgl1}), as a check on the Kubo
formula
calculation.  Thus for general $k$ in zero magnetic field and to order $k^2$ in
a magnetic field we have confirmed that the fluctuation contribution to
the dc conductivity matrix is indeed symmetric and given by (\ref{kubo}).
Note that the Hall conductivity is zero.\cite{ullah91}

The Gaussian approximation is valid in a region well above (as defined
by the Ginzburg criterion) the superconducting transition, where there
are only small amplitude fluctuations of the order parameter.  When
fluctuations are large, the nonlinear terms become important and the
Gaussian approximation no longer applies.  It is precisely in this
strong-fluctuation regime where a description in terms of fluctuating vortices
becomes appropriate.  This is the vortex liquid regime,
for which this calculation is inappropriate.  However, the low
temperature end of these Gaussian results may exhibit some of the
properties of the high temperature behavior of the technically more
difficult vortex liquid regime.
In principle, we could use the
Hartree approximation \cite{ullah91} to include a partial contribution
from the nonlinear terms.  However, because the Hartree approximation
only renormalizes the critical temperature and thus will not change
the signs or any other qualitative properties of $S_{ijkl}$, in this
paper we will only focus on the Gaussian approximation.

\section{Zero magnetic field}
\label{zerofield}

Before tackling the technically difficult task of calculating the
nonlocal conductivity tensor in an applied magnetic field,  we will
first calculate the nonlocal conductivity in zero magnetic field
in the Gaussian approximation.  In this case the system is translationally
invariant, and after Fourier transforming the Kubo formula,
Eq.\ (\ref{gaussian}), we obtain
\begin{equation}
\sigma_{\mu\nu}({\bf k})={2\over k_{B} T}\left({\hbar e^{*}\over 2m}\right)^{2}
\int {d\omega\over 2\pi} \int {d^{d}p\over (2\pi)^{d}} p_{\mu} p_{\nu}
 C_{0}({\bf p}+{\bf k}/ 2,\omega) C_{0}({\bf p}-{\bf k}/ 2,\omega),
\label{zerofield1}
\end{equation}
where the correlation function is
\begin{equation}
C_{0}({\bf k},\omega) = { 2 k_{B} T \Gamma^{-1} \over (\omega/\Gamma)^{2}
  + \left(\hbar^{2}k^{2}/ 2m + a\right)^{2} } .
\label{zerofield2}
\end{equation}
For $d<4$ the integral is ultraviolet-convergent, so we do not need a cutoff.
After substituting the correlation function into Eq.\ (\ref{zerofield1}),
performing the frequency integral, and scaling the momenta by the
correlation length $\xi = \hbar/\sqrt{2m|a|}$, we obtain
\begin{equation}
\sigma_{\mu\nu}({\bf k}) = \sigma(0) F_{\mu\nu}({\bf k}\xi),
\label{zerofield3}
\end{equation}
where the scaling function $F_{\mu\nu}(x)$ is normalized so that
$F_{\mu\nu}(0)=\delta_{\mu\nu}$, and where the
${\bf k}=0$ conductivity is \cite{aslamazov68,dorsey91}
\begin{equation}
\sigma(0) =  k_{B} T \left( { m (e^{*})^{2}\over \hbar^{4} \Gamma}\right)
 {\Gamma(2-d/2)\over (4\pi)^{d/2}}\xi^{4-d}.
\label{sigma0}
\end{equation}

The calculation of the scaling functions $F_{\mu\nu}({\bf k}\xi)$ is
rather complicated, and the details are relegated to Appendix A.
The conductivity can be decomposed into transverse and longitudinal components
\begin{equation}
\sigma_{\mu\nu}({\bf k}) =\sigma(0)[ F^{T}(k\xi) P^{T}_{\mu\nu}
                          + F^{L}(k\xi) P^{L}_{\mu\nu}],
\label{zerofield4}
\end{equation}
where we have introduced the transverse and longitudinal projection operators:
\begin{equation}
 P^{T} = \delta_{\mu\nu} - {k_{\mu}k_{\nu} \over k^{2}},\qquad
P^{L} = {k_{\mu}k_{\nu} \over k^{2}}.
\label{projection}
\end{equation}
A steady-state electric field is purely longitudinal (it is the gradient of the
scalar potential), so only the longitudinal part of the conductivity enters in
determining a dc current pattern.
In the Gaussian approximation, the transverse and longitudinal scaling
functions can be obtained in closed form (see Appendix A) and are plotted in
Figs. 2 and 3.  Both functions decrease monotonically
with increasing $k$.   As shown in Appendix A, for large $x$,
$F^{T,L}(x)\sim c_{d}^{T,L} (x/2)^{-(4-d)}$, with $c_{d}^{T,L}$
universal constants (there are logarithmic corrections in two dimensions).
Expanding the scaling functions to $O(k^{2})$, we obtain
\begin{eqnarray}
\sigma(0)F^{T,L}(k\xi) = \sigma(0) + S^{T,L}k^2 + ...  ,
\end{eqnarray}
with
\begin{equation}
S^{T} = -{5\over 48} (4-d) \sigma(0)\xi^{2}, \qquad
S^{L} = -{1\over 16} (4-d) \sigma(0)\xi^{2}.
\label{viscosity1}
\end{equation}
The full $S$ tensor can be written compactly as
\begin{equation}
S_{\mu\alpha\beta\nu} = S^{T}\delta_{\mu\nu}\delta_{\alpha\beta}
 + {1\over 2} (S^{L}-S^{T}) (\delta_{\mu\alpha}\delta_{\nu\beta}
        + \delta_{\mu\beta} \delta_{\nu\alpha}).
\label{vtensor}
\end{equation}
Finally, after Fourier transforming back to real space,  we
find for the current
\begin{equation}
{\bf J}({\bf x}) = \sigma(0){\bf E}({\bf x})
  + S^{T}\nabla\times\nabla\times {\bf E}({\bf x})
  + S^{L} \nabla \nabla\cdot {\bf E}({\bf x}) + ... .
\end{equation}

Note that $S^L$ (which is equal to, e.g., $S_{xxxx}$) is negative here in the
Gaussian approximation, and is argued above (Sec. III) to remain
negative in the vortex liquid in the critical regime
just above $T_c$ at $H=0$.  The general scaling form (5.5) presumably also
remains
valid in the critical regime, but with quantitatively different scaling
functions.  However, the qualitative behavior of the conductivity---that
it is maximal for $k=0$ and falls
off as a power of $k$ for large $k\xi$---should be the same in the nontrivial
critical
regimes for $H=0$.  Thus
there is no sign here that anything dramatic occurs in the nonlocal
conductivity at the Ginzburg crossover from mean-field to nontrivial
critical behavior in zero magnetic field.

\section{Nonzero magnetic field}

In this section we will calculate the nonlocal conductivity
in the Gaussian approximation in a uniform applied
magnetic field ${\bf H}= H \hat{z}$.  We will work in the Landau
gauge ${\bf A} = (0, Hx, 0)$.  The response function is
obtained as an eigenfunction expansion in a harmonic oscillator
basis set in the $x$ direction---the Landau-level expansion.
Using Eq. (\ref{correlation2}), we then obtain for the correlation
function in a mixed representation
\begin{equation}
C_{0}(x,x';k_{y},k_{z},\omega) = 2k_{B} T \Gamma^{-1}
\sum_{n=0}^{\infty} {u_{n}(x-x_{0}) u_{n}(x'-x_{0}) \over
 (\omega/\Gamma)^{2} + \varepsilon_{n,k_{z}}^{2}},
\label{magfield1}
\end{equation}
where the oscillator functions are
\begin{equation}
u_{n}(x) = \left(1\over 2^{n} n! \sqrt{\pi} l_{H}\right)^{1/2}
   e^{-x^{2}/2l_{H}^{2}} H_{n}(x/l_{H}),
\label{eigenstates}
\end{equation}
with the energies
\begin{equation}
\varepsilon_{n,k_{z}}= {\hbar^{2} k_{z}^{2}\over 2m}
 + \hbar\omega_{c}(n+{1\over2}) + a .
\label{eigenvalues}
\end{equation}
In the above we have introduced the magnetic length
$l_{H}=(\hbar/e^{*} H)^{1/2}$, the orbit center coordinate
$x_{0}=l_{H}^{2}k_{y}$ and the cyclotron frequency
$\omega_{c} = e^{*} H/m$.

\subsection{ ${\bf J}$, ${\bf E}$ perpendicular to ${\bf H}$}

In this subsection we calculate the nonlocal conductivity when
the current and electric field are in the $x-y$ plane.
Given the symmetries of the system, in this geometry at
order $k^2$ there are only three independent coefficients
to be calculated: $S_{yyyy}$, $S_{yxxy}$, and $S_{yzzy}$.
To obtain these it is only necessary to calculate $\sigma_{yy}({\bf k})$.
Rotational symmetry about the $z$-axis gives $S_{xxxx}=S_{yyyy}$,
$S_{xyyx}=S_{yxxy}$, $S_{xzzx}=S_{yzzy}$, etc. and
$S_{xxyy}=S_{yyxx}=(S_{yyyy}-S_{yxxy})/2$.  Due to the symmetries
and the absence of a Hall effect, all $S_{ijkl}$ with unpaired
indices (e.g., $S_{xyyy}$) vanish in this calculation.

To calculate $\sigma_{yy}({\bf k})$, we   carry out the
differentiations indicated in Eq.\ (\ref{gaussian}),
(noting that $D_{y}= \partial_{y} - ieHx/\hbar$) using the
correlation function in Eq.\ (\ref{magfield1}).  The frequency
integral is then easily performed; after scaling $x$ and $x_{0}$
by the magnetic length $l_{H}$,  and
rescaling the external momenta in the $x-y$ plane by
$l_{H}$ so that $(\bar{k}_{x},\bar{k}_{y})=(k_{x}l_{H},k_{y}l_{H})$,
we obtain
\begin{eqnarray}
 \sigma_{yy}({\bf k}) & = &  k_{B} T\Gamma^{-1} (e^{*})^{2}
\left( {e^{*} H\over m}\right)^{2} {1\over 2\pi}
\sum_{m,n=0}^{\infty} I_{mn}(\bar{k}_{x},\bar{k}_{y})\nonumber \\
  & & \times \int_{-\infty}^{\infty} {dp_{z}\over 2\pi}
 {1 \over \varepsilon_{m}(p_{z}+k_{z}/2)\varepsilon_{n}(p_{z}-k_{z}/2)
 [\varepsilon_{m}(p_{z}+k_{z}/2) + \varepsilon_{n}(p_{z}-k_{z}/2) ]},
\label{sigmayy1}
\end{eqnarray}
where
\begin{eqnarray}
 I_{mn}(\bar{k}_{x},\bar{k}_{y}) &=& \int_{-\infty}^{\infty} d(x_{1}-x_{2})
e^{i\bar{k}_{x}(x_{1}-x_{2})} \int_{-\infty}^{\infty} dx_{0}\, (x_{0}-x_{1})
 (x_{0}-x_{2})  u_{n}(x_{2}-x_{0}-\bar{k}_{y}/2)  \nonumber \\
 & & \times   u_{n}(x_{1}-x_{0}-\bar{k}_{y}/2)
  u_{m}(x_{1}-x_{0}+\bar{k}_{y}/2)  u_{m}(x_{2}-x_{0}+\bar{k}_{y}/2).
 \label{Imn}
\end{eqnarray}
After shifting the integration variables, this integral can be transformed
into a single integral:
\begin{equation}
I_{mn}(\bar{k}_{x},\bar{k}_{y}) = |A_{mn}^{(1)} (\bar{k}_{x},\bar{k}_{y})|^{2},
\label{Imn2}
\end{equation}
where
\begin{equation}
 A_{mn}^{(1)}(\bar{k}_{x},\bar{k}_{y}) = \int_{-\infty}^{\infty} dy\,
 e^{-i\bar{k}_{x}y} y \,  u_{m}(y+\bar{k}_{y}/2)  u_{n}(y-\bar{k}_{y}/2).
\label{Amn}
\end{equation}
Some simplification also occurs in the $p_{z}$-integral, if we first
rescale $p_{z}$ by the zero-temperature correlation length
$\xi(0)=(\hbar^{2}/2ma_{0})^{1/2}$, and introduce $\tilde{k}_{z} =
k_{z}\xi(0)$.
Then we find
\begin{equation}
\sigma_{yy}({\bf k}) =  {m\xi(0)  \over 8\pi\hbar^{2}\Gamma \Lambda_{T}}
  \sum_{m,n=0}^{\infty} |A_{mn}^{(1)}(\bar{k}_{x},\bar{k}_{y})|^{2}
 B_{mn}^{(1)}(\tilde{k}_{z}),
\label{newsigmayy}
\end{equation}
where we have introduced the thermal length
$\Lambda_{T}=\phi_{0}^{2}/16\pi^{2}k_{B}T$, \cite{fisher91}
the flux quantum $\phi_{0}=2\pi\hbar/e^{*}$,
and the scaled magnetic field $h=H/H_{c2}(0)$, with
$H_{c2}(0)=\phi_{0}/2\pi\xi^{2}(0)$ the zero temperature critical field.
The integral $B_{mn}^{(1)}$ is given by
\begin{equation}
B_{mn}^{(1)}(\tilde{k}_{z}) = 4 h^{2} \int_{-\infty}^{\infty} {dp\over 2\pi}
{1\over [(p + \tilde{k}_{z}/2)^{2} + \mu_{m}]
[(p -\tilde{k}_{z}/2)^{2} + \mu_{n}]
[p^{2} +(\tilde{k}_{z}/2)^{2} + \mu_{(m+n)/2}]},
\label{Bmn}
\end{equation}
where
\begin{equation}
\mu_{m} = \epsilon_{H} + 2hm, \qquad \epsilon_{H}= (T/T_{c})-1 + h.
\label{mu}
\end{equation}
Within mean-field theory the transition to the flux lattice state
occurs at $\epsilon_{H}=0$.

We now perform a long wavelength expansion of the conductivity.  The expansion
of $A_{mn}^{(1)}$ is most easily carried out by first noting that
it can be written in a compact operator form
(with $\hat{p}={1\over i} {\partial\over\partial_{y}}$ the momentum
operator),
\begin{eqnarray}
A_{mn}^{(1)} (\bar{k}_{x},\bar{k}_{y}) & = &  \langle m|
            e^{-{i\over 2}\bar{k}_{y}\hat{p}}
e^{-{i\over 2}\bar{k}_{x} \hat{y}} \hat{y} e^{-{i\over 2}\bar{k}_{x} \hat{y}}
 e^{-{i\over 2}\bar{k}_{y}\hat{p}} |n\rangle\nonumber \\
& = &e^{{i\over 2}\bar{k}_{x}\bar{k}_{y}}
\langle m| e^{-{i\over 2}( \bar{k}_{x} \hat{y}+ \bar{k}_{y}\hat{p})}
 \hat{y}  e^{-{i\over 2}( \bar{k}_{x} \hat{y}+ \bar{k}_{y}\hat{p})}|n\rangle,
\label{newAmn}
\end{eqnarray}
where we have used $[\hat{y},\hat{p}]=i$ to obtain the last line.
Expanding for small $(\bar{k}_{x},\bar{k}_{y})$, we find (neglecting the
multiplicative phase factor)
\begin{eqnarray}
A_{mn}^{(1)}(\bar{k}_{x},\bar{k}_{y}) & = &  \langle m|\hat{y}|n\rangle
 - i\langle m|\hat{y}^{2}|n\rangle \bar{k}_{x}
 - {i\over 2}\langle m|\hat{y}\hat{p}+ \hat{p}\hat{y}|n\rangle\bar{k}_{y}
   -{1\over 2} \langle m|\hat{y}^{3}|n\rangle \bar{k}_{x}^{2}
           \nonumber \\
& & \quad -{1\over 8}\langle m|\hat{y}\hat{p}^{2} + 2\hat{p}\hat{y}\hat{p}
       +  \hat{p}^{2}\hat{y} | n\rangle \bar{k}_{y}^{2}
	   \nonumber \\
& & \quad -{1\over 8} \langle m|3\hat{y}^{2}\hat{p} + 2\hat{y}\hat{p}\hat{y}
      + 3\hat{p}\hat{y}^{2} |n\rangle \bar{k}_{x}\bar{k}_{y} + O(k^3) .
\label{expandA}
\end{eqnarray}
The oscillator matrix elements are tabulated in Appendix B.
After squaring, we find
\begin{eqnarray}
|A_{mn}^{(1)}(\bar{k}_{x},\bar{k}_{y})|^{2} & = & {1\over
2}\left[n\,\delta_{m,n-1}
+ (n+1)\,\delta_{m,n+1}\right]  + \left[ (n-1)n\,\delta_{m,n-2}
 - 3n^{2}\,\delta_{m,n-1} \right.\nonumber \\
& &  \left.+(2n+1)^{2}\,\delta_{m,n} - 3(n+1)^{2}\,\delta_{m,n+1}
 + (n+1)(n+2)\,\delta_{m,n+2}\right] (\bar{k}_{x}/2)^{2} \nonumber \\
& &  + \left[ (n-1)n\,\delta_{m,n-2}
 - n^{2}\,\delta_{m,n-1} -(n+1)^{2}\,\delta_{m,n+1} \right.\nonumber\\
& &\left.   + (n+1)(n+2)\,\delta_{m,n+2}\right]
   (\bar{k}_{y}/2)^{2}  + O(\bar{k}_{x}^{4},\bar{k}_{y}^{4}).
\label{expandA2}
\end{eqnarray}
The long wavelength expansion of $B_{mn}^{(1)}(\tilde{k}_{z})$ is discussed in
Appendix B.
After substituting these expansions, Eqs.\ (\ref{expandA2}) and
(\ref{expandB2})  into the  expression for the conductivity,
Eq.\ (\ref{newsigmayy}), and returning to conventional units, we
obtain the ${\bf k}=0$ conductivity
\begin{equation}
\sigma_{yy}(0) =  {m\xi(0)  \over 8\pi\hbar^{2}\Gamma \Lambda_{T}h^{1/2}}
  \sum_{n=0}^{\infty}(n+1)\left[{1\over (\alpha+2n)^{1/2}}
- {2\over (\alpha+2n+1)^{1/2}} + {1\over (\alpha+2n+2)^{1/2}}\right],
\label{sigmayy0}
\end{equation}
which agrees with previous results.\cite{schmidt68,usadel69,mikeska70,ullah91}
We have defined  the scaling variable
$\alpha = \epsilon_H/h$, which measures the temperature
distance from from the mean-field $H_{c2}$ line in units proportional
to the magnetic field.  Thus $\alpha$ runs from zero on the $H_{c2}$
line at nonzero field to infinity at $H=0$ in the normal state above $T_c$.
Note that the sum here (and in many of the results below) is a scaling
function that depends only on this one parameter, $\alpha$. This type of sum
may be written in a more compact form by using the integral representation
for the function $\Phi (z,s,v)$. \cite{integraltable}  After resummation, they
are the {\it Laplace transformations} of certain functions, so we define
\begin{eqnarray}
\sigma_{ij} (0) &=& \sigma(0) \, (\alpha-1)^{1/2} \, \int_0^{\infty} \,
\frac{dt
}{\sqrt \pi}  \, \exp (-\alpha t) \,\, \tilde{\sigma}_{ij}(0) \, ,
\label{laplace1} \\
S_{ijkl} &=& \sigma(0) \, \xi^{2}(T) \,  (\alpha-1)^{3/2} \, \int_0^{\infty}
\,
\frac{dt}{\sqrt \pi} \, \exp (-\alpha t) \,\, \tilde{S}_{ijkl} \,
,\label{laplace2}
\end{eqnarray}
where we have expressed the prefactor
$m\xi(0) /  8\pi\hbar^{2}\Gamma \Lambda_{T}$ in terms of $\alpha$,
the zero-field coherence length $\xi(T)$, and $\sigma (0)$, the $k=0$, $H=0$
conductivity at temperature $T$.  We find
\begin{eqnarray}
\tilde{\sigma}_{yy}(0) = \frac{4 t^{-1/2 }}{(1+\exp (-t))^2} \, .
\label{sigmayy2}
\end{eqnarray}

 We also obtain:
\begin{eqnarray}
S_{yyyy} & =&  {m\xi(0)  \over 8\pi\hbar^{2}\Gamma \Lambda_{T} h^{1/2}}
\left({l_{H}\over 2}\right)^{2} \sum_{n=0}^{\infty}\left[-{(n+1)(3n+2)\over
2(\alpha+2n)^{1/2}} + {4(n+1)^{2}\over (\alpha+2n+1)^{1/2}}\right. \nonumber \\
& & \qquad \left. - {(n+1)(3n+4)\over (\alpha+2n+2)^{1/2}}
    + {(n+1)(n+2)\over 2(\alpha+2n+4)^{1/2}} \right] \, , \nonumber \\
& & \nonumber \\
\tilde{S}_{yyyy} &=& \frac{- t^{-1/2 } (1- \exp (-t))}{(1+\exp (-t))^3}  \, ,
\label{Syyyy}
\end{eqnarray}
\begin{eqnarray}
S_{yxxy} &=& {m\xi(0)  \over 8\pi\hbar^{2}\Gamma \Lambda_{T}h^{1/2}}
\left({l_{H}\over 2}\right)^{2} \sum_{n=0}^{\infty}\left[ {3\over 4}
{(2n+1)^{2}\over (\alpha+2n)^{5/2}} - {(n+1)(11n+10)\over 2(\alpha+2n)^{1/2}}
 \right. \nonumber \\
& & \quad \left. + {12 (n+1)^{2}\over (\alpha+2n+1)^{1/2} }
- {(n+1)(7n+8)\over (\alpha+2n+2)^{1/2}} + {(n+1)(n+2)\over
2(\alpha+2n+4)^{1/2}}
\right] \, , \nonumber \\
& & \nonumber \\
\tilde{S}_{yxxy} &=& t^{-1/2} \left[ \frac{(t^2-5) \exp(-4 t) + 12 \exp(-3t)}
{(1- \exp (-2t)) ^3} \right. \nonumber \\
&+& \left. \frac{(6 t^2-1 4) \exp(-2t) + 12 \exp(-t)
+t^2-5}{(1- \exp (-2t)) ^3} \right] \, , \nonumber \\
& &
\label{Syxxy}
\end{eqnarray}

\begin{eqnarray}
S_{yzzy} & =&  {m\xi(0)  \over 8\pi\hbar^{2}\Gamma \Lambda_{T}h^{3/2}}
\left(\xi(0)\over2\right)^{2} \sum_{n=0}^{\infty} \left\{
{(n+1)\over (\alpha+2n+1)^{3/2}} \right.\nonumber \\
 & & \quad   + 4(n+1)\left[ {1\over (\alpha+2n+2)^{1/2}}
 - {1\over (\alpha+2n)^{1/2}}\right]  \nonumber \\
 & & \qquad  - 12(n+1)\left((\alpha+2n)^{1/2} - 2(\alpha+2n+1)^{1/2}
 + (\alpha+2n+2)^{1/2}\right) \Biggr\} \, , \nonumber \\
\tilde{S}_{yzzy} &=& 2 t^{-3/2} \left[ (2 t+3) \exp(-2t) + (t^2-6) \exp(-t)
+3-2t
 \right] \over (1- \exp (-2t))^2 \, .
\label{Syzzy}
\end{eqnarray}

An advantage of the integral representations is that
one can determine the signs much more easily.  We find
that for $t>0$, $\tilde{S}_{ijkl}$ and $\tilde{\sigma}_{ij}$ are
either always positive ($\tilde{S}_{yyzz}$, $\tilde{\sigma}_{yy}(0)$,
and $\tilde{\sigma}_{zz}(0)$) or always negative
($\tilde{S}_{yyyy}$, $\tilde{S}_{zzzz}$, $\tilde{S}_{zyyz}$,
$\tilde{S}_{yzzy}$ , $\tilde{S}_{zyyz} + 2 \tilde{S}_{zzyy}$,
and $\tilde{S}_{yzzy} + 2 \tilde{S}_{yyzz}$).
The only exception is $\tilde{S}_{yxxy}$ which changes sign at about t=1.45.
As a result, $S_{yxxy}$ changes sign at about $\alpha=1.2$ ($\tilde{S}_{yxxy}$
becomes positive for $\alpha \approx 1.2$).

The zero field limit ($\alpha \rightarrow \infty$) can also
be easily obtained in the integral representations. Simply by redefining
 $\alpha t $ as $ x$ in (\ref{laplace1}) and (\ref{laplace2})
 and then keeping only the leading order terms
of the power series in $x$ of the integrands, we can recover
the zero field results of the previous section.

\subsection{{\bf J}, {\bf E} parallel to {\bf H}}

In this geometry we calculate $\sigma_{zz}({\bf k})$.
The calculation of $\sigma_{zz}$ closely parallels the calculation
of $\sigma_{yy}$, so we will only outline the results.  First, we
can express the conductivity as
\begin{equation}
\sigma_{zz}({\bf k}) =  {m\xi(0)  \over 32\pi\hbar^{2}\Gamma \Lambda_{T}} h
  \sum_{m,n=0}^{\infty} |A_{mn}^{(2)}(\bar{k}_{x},\bar{k}_{y})|^{2}
 B_{mn}^{(2)}(\tilde{k}_{z}),
\label{newsigmazz}
\end{equation}
where
\begin{eqnarray}
A_{mn}^{(2)}(\bar{k}_{x},\bar{k}_{y}) & = &  \int_{-\infty}^{\infty} dy\,
 e^{-i\bar{k}_{x} y} u_{m}(y+\bar{k}_{y}/2) u_{n}(y-\bar{k}_{y}/2)\nonumber \\
  &=& e^{{i\over 2}\bar{k}_{x}\bar{k}_{y}}
  \langle m| e^{i( \bar{k}_{x} \hat{y}+ \bar{k}_{y}\hat{p})} |n\rangle,
\label{Amn2}
\end{eqnarray}
and
\begin{equation}
B_{mn}^{(2)}(\tilde{k}_{z}) = 16  \int_{-\infty}^{\infty} {dp\over 2\pi}
{p^{2} \over [(p +\tilde{k}_{z}/2)^{2} + \mu_{m}]
[(p -\tilde{k}_{z}/2)^{2} + \mu_{n}]
[p^{2} +(\tilde{k}_{z}/2)^{2} + \mu_{(m+n)/2}]}.
\label{Bmn2}
\end{equation}
Performing the long wavelength expansion with the help of the results
in Appendix B, we have  for $A_{mn}^{(2)}$
\begin{eqnarray}
|A_{mn}^{(2)}(\bar{k}_{x},\bar{k}_{y})|^{2} & =&  \delta_{m,n}
 +{1\over 2} \left[ n\delta_{m,n-1} - (2n+1)\delta_{m,n}
       +(n+1)\delta_{m,n+1}\right] (\bar{k}_{x}^{2} + \bar{k}_{y}^{2})\nonumber
\\
& & +O(\bar{k}_{x}^{4},\bar{k}_{y}^{4},\bar{k}_{x}^{2}\bar{k}_{y}^{2}).
\label{expandAmn2}
\end{eqnarray}
Combining this with the long wavelength expansion of $B_{mn}^{(2)}$,
we find for the conductivity
\begin{eqnarray}
\sigma_{zz}(0) & = &  {m\xi(0)  \over 32\pi\hbar^{2}\Gamma \Lambda_{T}h^{1/2}}
  \sum_{n=0}^{\infty}{1\over (\alpha+2n)^{3/2}} \, , \nonumber \\
\tilde{\sigma}_{zz}(0) & = & \frac{2 t^{1/2 }}{ 1- \exp (-2t)} \, ,
\label{sigmazz0}
\end{eqnarray}
which again agrees with previous results,
\cite{schmidt68,usadel69,mikeska70,ullah91} and
\begin{eqnarray}
S_{zyyz} &=& - {m\xi^{3}(0)  \over 32\pi\hbar^{2}\Gamma \Lambda_{T} h^{3/2}}
 \sum_{n=0}^{\infty} \left[ {n+1/2\over (\alpha+2n)^{3/2}} \right. \nonumber \\
& & \quad  + 4(n+1) \left( (\alpha+2n)^{1/2} - 2(\alpha+2n+1)^{1/2}
               + (\alpha+2n+2)^{1/2} \right) \Biggr] \, , \nonumber \\
& & \nonumber \\
\tilde{S}_{zyyz} &=& \frac{t^{-3/2} \left[ (t^2-2) \exp(-2t) +4 \exp(-t) +t^2-2
\right
]}{ (1- \exp (-2t))^2 } \, ,
\label{Szyyz}
\end{eqnarray}
\begin{eqnarray}
S_{zzzz} &=& - {3 m\xi^{3}(0)  \over 512 \pi\hbar^{2}\Gamma \Lambda_{T}h^{3/2}}
  \sum_{n=0}^{\infty}{1\over (\alpha+2n)^{5/2}} \, ,\nonumber \\
\tilde{S}_{zzzz} &=& \frac{- t^{3/2}}{ 4 (1- \exp (-2t))} \, .
\label{Szzzz}
\end{eqnarray}

\subsection{{\bf J} perpendicular to {\bf H}, {\bf E} parallel to {\bf H}}

We now need to calculate $\sigma_{yz}({\bf k})$.
Following the same steps as in the previous two sections,
we find for the conductivity
\begin{equation}
\sigma_{yz}({\bf k}) = - {m\xi(0)  \over 8\pi\hbar^{2}\Gamma \Lambda_{T}}
  {1\over h^{1/2}}
  \sum_{m,n=0}^{\infty} A_{mn}^{(1)}(\bar{k}_{x},\bar{k}_{y})^{*}
                         A_{mn}^{(2)}(\bar{k}_{x},\bar{k}_{y})
                       B_{mn}^{(3)}(\tilde{k}_{z}),
\label{sigmayz}
\end{equation}
where
\begin{equation}
 B_{mn}^{(3)}(\tilde{k}_{z}) = 4 h^{2}  \int_{-\infty}^{\infty} {dp\over 2\pi}
{p \over [(p +\tilde{k}_{z}/2)^{2} + \mu_{m}]
[(p -\tilde{k}_{z}/2)^{2} + \mu_{n}]
[p^{2} +(\tilde{k}_{z}/2)^{2} + \mu_{(m+n)/2}]}.
\label{Bmn3}
\end{equation}
Expanding the $A_{mn}$'s using the results of the previous sections, we find
\begin{eqnarray}
 A_{mn}^{(1)}(\bar{k}_{x},\bar{k}_{y})^{*}
A_{mn}^{(2)}(\bar{k}_{x},\bar{k}_{y})
 &=& i[n\delta_{m,n-1} + (2n+1)\delta_{m,n} +
(n+1)\delta_{m,n+1}](\bar{k}_{x}/2)
   \nonumber \\
 & & \qquad - [ -n\delta_{m,n-1} + (n+1)\delta_{m,n+1}](\bar{k}_{y}/2)
  + O(\bar{k}_{x}^{2} ,\bar{k}_{y}^{2},\bar{k}_{x}\bar{k}_{y}).
\label{expandAmn3}
\end{eqnarray}
Combining this result with the long wavelength expansion of $B_{mn}^{(3)}$
in Appendix B, we obtain
\begin{eqnarray}
S_{yyzz} & =&  {m\xi^{3}(0)\over 32 \pi\hbar^{2}\Gamma \Lambda_{T}}{1\over h}
 \sum_{n=0}^{\infty}\left[ (n+1)\left({1\over \mu_{n}^{1/2}}
  - {1\over \mu_{n+1}^{1/2}}\right)\right. \nonumber \\
 & & \qquad\qquad \left. +{4(n+1)\over h} \left(\mu_{n}^{1/2} + \mu_{n+1}^{1/2}
      -2\mu_{n+1/2}^{1/2}\right)\right] \, , \nonumber \\
\tilde{S}_{yyzz} &=&  t^{-3/2} \left[ (t+2) \exp(-t) + t-2 \right] \over (1-
\exp
 (-2t)) ((1+\exp (-t)) \, .
\label{Syyzz}
\end{eqnarray}
In zero field, $S_{yyzz}$ approaches $(S^L-S^T)/2$ which is positive.

Finally, we find that
$\sigma_{zy}(k_{x},k_{y},k_{z})=\sigma_{yz}(-k_{x},k_{y},k_{z})$,
so that $S_{zzyy}=S_{yyzz}$.

\subsection{General geometry with longitudinal electric field}

In a truly dc steady state the electric field must be purely
longitudinal.  Thus let us consider general wavevector ${\bf k}$
and ask what the current is in linear response to a longitudinal
electric field.  Without loss of generality, we can take
${\bf k}=k_y{\bf \hat y} + k_z{\bf \hat z}$, $E_y=Ek_y/k$ and
$E_z=Ek_z/k$.  To order $k^2$, the resulting current is
\begin{eqnarray}
J_y & = & Ek_y[\sigma_{yy}(0)+S_{yyyy}k_y^2+(S_{yzzy}+2S_{yyzz})k_z^2]/k ,
\label{jy} \\
J_z & = & Ek_z[\sigma_{zz}(0)+S_{zzzz}k_z^2+(S_{zyyz}+2S_{zzyy})k_y^2]/k ,
\label{jz}
\end{eqnarray}
and, in the absence of a Hall effect, $J_x=0$.  Thus we see that
in this geometry, the nonlocal effect here in the Gaussian approximation
is always of the sign such that the conductivity for a longitudinal
electric field is reduced as ${\bf k}$ moves away from {\bf 0}.
This is of the opposite sign from what we argue above occurs in the
vortex liquid regime at lower temperatures.

\section{Discussion and Conclusions}

In this paper we have examined the wavevector-dependent
dc conductivity, $\sigma({\bf k})$, of
a type-II superconductor in various regimes, using phenomenological arguments
and the TDGL equation.  There appear to be at least four
qualitatively different regimes of behavior for $\sigma({\bf k})$:
First, in the Meissner phase the conductivity is infinite at $k=0$
and monotonically decreasing with increasing $k$.  This behavior
should also apply in the pinned vortex lattice and vortex glass
phases, where there is no vortex motion in linear response to a
uniform applied current.  Second, in an ideal, unpinned vortex
lattice phase, the conductivity is
discontinuous at $k=0$, taking on the finite, flux-flow value at
$k=0$ due to vortex motion, but varying as $k^{-2}$ for small,
positive $k$, where there is no vortex motion.  Here the conductivity
is still a monotonically decreasing function of $k$ for $k>0$, but is now
non-monotonic when $k=0$ is included.  Third, for all temperatures
above $T_c$  in zero magnetic field and for sufficiently high
temperatures in small nonzero magnetic fields, the qualitative
behavior seen in the above Gaussian-order TDGL calculation applies.
There the conductivity is finite and maximal for $k=0$
and is smooth and monotonically decreasing with increasing $k$.  As argued on
a phenomenological level in Section III above, this behavior should
apply in the vortex liquid in a low-field regime where the dissipation is
dominated by the
spontaneous, thermally-excited vortices rather than the field-induced
vortices.  Last, phenomenological arguments suggest that in the vortex
liquid regime at sufficiently low temperatures and high fields the conductivity
is instead a non-monotonic function of $k$:  At small $k$ the conductivity
increases with $k$ due to the large vortex-liquid viscosity that
impedes nonuniform motion of the vortex liquid.  However, at larger $k$,
where vortex motion is not the dominant effect in determining the conductivity,
the more microscopic behavior of a conductivity that decreases with
increasing $k$ prevails, as it does at large $k$ in all regimes.
It is this last vortex liquid regime that is the least accessible
theoretically,
because it is a strong thermal fluctuation regime that does not
exist in a mean-field or weak-fluctuation treatment of Ginzburg-Landau theory.
Thus the qualitative behavior described for the first three regimes can be
obtained directly from the TDGL equations, while the theoretical support
for the description of the last, vortex liquid regime is,
at this time, purely phenomenological.

What other experiments might be done to probe the $k$-dependence of the
transport properties?  In a transport experiment one has access only
to the surface of the sample.  The voltage contacts can measure the
electric field parallel to the surface and the current contacts can
set $\nabla \cdot {\bf J}$ at the surface.  For a bulk sample, this
means one must rely on modeling to deduce what is going on inside
the sample.  However, in a film geometry, the entire sample is surface,
so one could, in principle, have much more complete measurements and
control.  With modern microfabrication techniques, it seems a study
that probes down to micron or shorter length scales should be
feasible.  We pose this as a future experimental challenge.

\acknowledgments
C.-Y.M. and A.T.D. acknowledge support from NSF Grant DMR 92-23586
and A.T.D. gratefully acknowledges an Alfred P. Sloan Foundation
Fellowship.  R.W. gratefully acknowledges an AT\&T PhD Fellowship as
well as useful discussions with Anthony J. Leggett.

\appendix

\section{Zero field scaling function}

In this Appendix we will calculate the integrals which appear in the
scaling function for the conductivity in zero magnetic field.
Letting $\bar{\bf k}={\bf k}\xi $ be a dimensionless
momentum, we have for the scaling function in dimension $d$
\begin{eqnarray}
F_{\mu\nu} (\bar{\bf k})& =& {4 (4\pi)^{d/2}\over \Gamma(2-d/2)}
 \int {d^{d}p\over (2\pi)^{d}} p_{\mu} p_{\nu}\nonumber \\
 & & \qquad \times { 1 \over [p^{2} + (\bar{k}/2)^{2}
+ {\bf p}\cdot\bar{\bf k}+ 1][p^{2} + (\bar{k}/2)^{2}
- {\bf p}\cdot\bar{\bf k}+ 1] [p^{2} + (\bar{k}/2)^{2} + 1]}.
\label{F}
\end{eqnarray}
In order to simplify the integrals we use the Feynman parameterization
\cite{amit}  to first combine the first two terms in the denominator,
and then once again to fold in the third term, with the result that
\begin{eqnarray}
& & {1\over [p^{2} + (\bar{k}/2)^{2}
+ {\bf p}\cdot\bar{\bf k}+ 1][p^{2} + (\bar{k}/2)^{2}
- {\bf p}\cdot\bar{\bf k}+ 1] [p^{2} + (\bar{k}/2)^{2} + 1]}\nonumber \\
& & \qquad = 2 \int_{0}^{1}dx\int_{0}^{1}dy\,{y\over [p^{2}+(\bar{k}/2)^{2} + 1
       + (2x-1)y\, {\bf p}\cdot\bar{\bf k}]^{3}}.
\label{feynman}
\end{eqnarray}
We then substitute this result into Eq.\ (\ref{F}), change variables in the
momentum integral to ${\bf q} = {\bf p} + (2x-1)y (\bar{\bf k}/2)$ to
eliminate the
terms linear in $\bar{\bf k}$, and perform the $d$-dimensional momentum
integral.\cite{amit}  We are then left with
\begin{equation}
F_{\mu\nu} (\bar{k}) = F^{T}(\bar{k}) P^{T}_{\mu\nu}
       + F^{L}(\bar{k}) {\bar{k}_{\mu}\bar{k}_{\nu} \over\bar{k}^{2}},
\label{newF}
\end{equation}
where
\begin{equation}
F^{T}(\bar{k}) = 2 \int_{0}^{1} dw\,\int_{0}^{1}dy\, {y\over
 [1 + (\bar{k}/2)^{2}(1-w^{2}y^{2})]^{2-d/2} },
\label{FT}
\end{equation}
\begin{equation}
F^{L}(\bar{k}) = F^{T}(\bar{k})+  (2-d/2)\bar{k}^{2}\int_{0}^{1} dw\,
 \int_{0}^{1} dy\, { w^{2}y^{3}\over
[1+(\bar{k}/2)^{2}(1-w^{2}y^{2})]^{3-d/2}},
\label{FL}
\end{equation}
and where we have changed variables to $w=2x-1$.  The integrals on $y$
can be performed, and the remaining integrals on $w$ can be  simplified
by integrating by parts.  We finally obtain
\begin{equation}
F^{T}(\bar{k}) =  2 \int_{0}^{1} dw\, [1 + (\bar{k}/2)^{2}(1-w^{2})]^{d/2-2}
- {2\over (d-2)}{ [1+(\bar{k}/2)^{2}]^{d/2-1}- 1 \over (\bar{k}/2)^{2}},
\label{finalFT}
\end{equation}
\begin{equation}
F^{L}(\bar{k}) = {2\over (d-2)}{ [1+(\bar{k}/2)^{2}]^{d/2-1}
- 1 \over (\bar{k}/2)^{2}}.
\label{finalFL}
\end{equation}
For $d=4$ we have $F^{T}=F^{L}=1$; for $d=3$ we have
\begin{equation}
F^{T}(\bar{k})= 2 { \sin^{-1}[(\bar{k}/2)/\sqrt{1+(\bar{k}/2)^{2}}]
 \over (\bar{k}/2)} - 2 {\sqrt{1 + (\bar{k}/2)^{2}} - 1 \over (\bar{k}/2)^{2}},
\label{d3FT}
\end{equation}
\begin{equation}
F^{L}(\bar{k}) = 2 {\sqrt{1 + (\bar{k}/2)^{2}} - 1 \over (\bar{k}/2)^{2}},
\label{d3FL}
\end{equation}
while for $d=2$
\begin{equation}
F^{T}(\bar{k}) = 2{\coth^{-1}[\sqrt{1+(\bar{k}/2)^{2}}/(\bar{k}/2)]
 \over (\bar{k}/2)\sqrt{1+(\bar{k}/2)^{2}}}
  - {\ln[1+(\bar{k}/2)^{2}]\over (\bar{k}/2)^{2}},
\label{d2FT}
\end{equation}
\begin{equation}
F^{L}(\bar{k}) = {\ln[1+(\bar{k}/2)^{2}]\over (\bar{k}/2)^{2}}.
\label{d2FL}
\end{equation}
Plots of these scaling functions are shown in Figs. 2 and 3.
For small $\bar{k}$ these expressions have the following expansions
for general dimension $d$:
\begin{equation}
F^{T}(\bar{k})= 1 - {5\over 48}(4-d)\bar{k}^{2} + O(\bar{k}^{4}),
\label{smallkFT}
\end{equation}
\begin{equation}
F^{L}(\bar{k}) = 1-{1\over 16}(4-d)\bar{k}^{2} + O(\bar{k}^{4}).
\label{smallkFL}
\end{equation}
For large $\bar{k}$ and $d\neq 2$, we have
\begin{equation}
F^{T}(\bar{k})\sim c^{T}_{d}(\bar{k}/2)^{-(4-d)},
\qquad F^{L}(\bar{k}^{2})\sim c^{L}_{d}(\bar{k}/2)^{-(4-d)},
\label{largek}
\end{equation}
where the constants are functions of dimension $d$, given by
\begin{equation}
c^{T}_{d}= 2\int_{0}^{1}dw\,(1-w^{2})^{d/2-2} - {2\over d-2},\qquad
c^{L}_{d} = {2\over d-2},
\label{constants}
\end{equation}
with $c^{T}_{d=3}= \pi -2$.  For large $\bar{k}$ and $d=2$,
\begin{equation}
F^{T}(\bar{k})\sim {2\ln 2 \over (\bar{k}/2)^{2}},\qquad
F^{L}(\bar{k}) \sim 2 {\ln (k/2) \over (k/2)^{2}}.
\label{d2constants}
\end{equation}
The large $\bar{k}$ behavior agrees with the result of the scaling theory
discussed in Sec. III, up to some logarithmic corrections in two
dimensions.

\section{Expansions for integrals in a magnetic field}

In this Appendix we will include some of the details of the calculation of the
viscosities in a magnetic field.  First, we simply list some of the
harmonic oscillator matrix elements which are used to evaluate the
integrals $A_{mn}^{(1)}$ and $A_{mn}^{(2)}$:
\begin{equation}
\langle m | \hat{y} | n \rangle = {1\over \sqrt{2}}\left[
\sqrt{n}\, \delta_{m,n-1} +   \sqrt{n+1}\, \delta_{m,n+1}\right] ,
\label{osc1}
\end{equation}
\begin{equation}
\langle m | \hat{y}^{2} | n \rangle = {1\over 2}\left[
\sqrt{(n-1)n}\, \delta_{m,n-2} +(2 n+ 1)\, \delta_{m,n}
 + \sqrt{(n+1)(n+2)}\, \delta_{m,n+2}\right] ,
\label{osc2}
\end{equation}
\begin{eqnarray}
\langle m | \hat{y}^{3} | n \rangle &=&  {1\over 2\sqrt{2}}\left[
\sqrt{(n-2)(n-1)n}\, \delta_{m,n-3} + 3n^{3/2}\,\delta_{m,n-1}\right.
         \nonumber \\
& & \qquad+ \left. 3 (n+1)^{3/2}\,\delta_{m,n+1}
+ \sqrt{(n+1)(n+2)(n+3)}\, \delta_{m,n+3}\right],
\label{osc3}
\end{eqnarray}
\begin{equation}
\langle m | \hat{p} | n \rangle ={i\over \sqrt{2}}\left[
-\sqrt{n}\, \delta_{m,n-1}  + \sqrt{n+1}\, \delta_{m,n+1} \right],
\label{osc4}
\end{equation}
\begin{equation}
\langle m  | \hat{p}^{2} | n \rangle = {1\over 2} \left[
-\sqrt{(n-1)n}\, \delta_{m,n-2} + (2n+1)\,\delta_{m,n}
  - \sqrt{(n+1)(n+2)}\, \delta_{m,n+2}\right],
\label{osc5}
\end{equation}
\begin{equation}
\langle m | \hat{y}\hat{p} + \hat{p}\hat{y} |n\rangle =
i\left[-\sqrt{(n-1)n}\,\delta_{m,n-2}
  + \sqrt{(n+1)(n+2)}\,\delta_{m,n+2} \right],
\label{osc6}
\end{equation}
\begin{eqnarray}
\langle m | \hat{y}\hat{p}^{2} + 2\hat{p}\hat{y}\hat{p} + \hat{p}^{2}\hat{y}
  |n\rangle  & =& -{2\over \sqrt{2}} \left[\sqrt{(n-2)(n-1)n}\,\delta_{m,n-3}
  - n^{3/2}\,\delta_{m,n-1} \right. \nonumber \\
& & \quad \left. -(n+1)^{3/2}\,\delta_{m,n+1}
         +\sqrt{(n+1)(n+2)(n+3)}\,\delta_{m,n+3}\right],
\label{osc7}
\end{eqnarray}
\begin{eqnarray}
\langle m | 3\hat{y}^{2}\hat{p} +2 \hat{y}\hat{p}\hat{y} + 3\hat{p}\hat{y}^{2}
  |n\rangle  & =& {4i\over \sqrt{2}} \left[-\sqrt{(n-2)(n-1)n}\,\delta_{m,n-3}
  - n^{3/2}\,\delta_{m,n-1} \right. \nonumber \\
& & \quad \left. +(n+1)^{3/2}\,\delta_{m,n+1}
         +\sqrt{(n+1)(n+2)(n+3)}\,\delta_{m,n+3}\right].
\label{osc8}
\end{eqnarray}

Next, we will consider the long wavelength expansion of the integrals
$B_{mn}^{(1)}$, $B_{mn}^{(2)}$, $B_{mn}^{(3)}$.  These can be efficiently
evaluated by using the Feynman parameterization, similar to what
was done in Appendix B.  We then have for $B_{mn}^{(1)}$
\begin{equation}
B_{mn}^{(1)}(\tilde{k}_{z}) = {3 h^{2}\over 4} \int_{-1}^{1} dw \int_{0}^{1}
dy\,
{y\over \left[ (\tilde{k}_{z}/2)^{2}(1-w^{2}y^{2}) + h(m-n)wy
          + \mu_{(m+n)/2}\right]^{5/2}}.
\label{expandB1}
\end{equation}
Expanding for small $\tilde{k}_{z}$, we find
\begin{eqnarray}
B_{mn}^{(1)}(\tilde{k}_{z}) & =&  {1\over (m-n)^{2}} \left[ {1\over
\mu_{m}^{1/2}}
 + {1\over \mu_{n}^{1/2}} - {2\over \mu_{(m+n)/2}^{1/2}}\right] \nonumber\\
 & & \quad + \left\{ {1\over (m-n)^{2}}{1\over \mu_{(m+n)/2}^{3/2}}
 + {1\over (m-n)^{3}}{4\over h} \left[{1\over \mu_{m}^{1/2}}
         -{1\over\mu_{n}^{1/2}}\right] \right.\nonumber \\
 & & \quad\quad \left. - {1\over (m-n)^{4}}{12\over h^{2}}\left[
\mu_{n}^{1/2} + \mu_{m}^{1/2} - 2
\mu_{(m+n)/2}^{1/2}\right]\right\}(\tilde{k}_{z}/2)^{2}
 + O(\tilde{k}_{z}^{4}) .
\label{expandB2}
\end{eqnarray}
In a similar fashion we have
\begin{eqnarray}
B_{mn}^{(2)}(\tilde{k}_{z}) &=&  \int_{-1}^{1} dw \int_{0}^{1} dy\,
{y\over \left[ (\tilde{k}_{z}/2)^{2}(1-w^{2}y^{2}) + h(m-n)wy
          + \mu_{(m+n)/2}\right]^{3/2}}\nonumber \\
& & \quad + 3 (\tilde{k}_{z}/2)^{2}  \int_{-1}^{1} dw \int_{0}^{1} dy\,
{w^{2}y^{3}\over \left[ (\tilde{k}_{z}/2)^{2}(1-w^{2}y^{2}) + h(m-n)wy
          + \mu_{(m+n)/2}\right]^{5/2}}\nonumber \\
 &=&  - {4\over h^{2} (m-n)^{2}} [\mu_{m}^{1/2}+\mu_{n}^{1/2}
 - 2\mu_{(m+n)/2}^{1/2}]  - {4\over h^{2} (m-n)^{2}}
     \left[{1\over \mu_{m}^{1/2}} + {1\over \mu_{n}^{1/2}}
            \right. \nonumber \\
 & &  \left.   + {1\over \mu_{(m+n)/2}^{1/2}}
     + {12 \mu_{(m+n)/2}\over h^{2}(m-n)^{2}} (\mu_{m}^{1/2}
    + \mu_{n}^{1/2}) -24 {\mu_{(m+n)/2}^{3/2}\over h^{2}(m-n)^{2}}\right]
 (\tilde{k}_{z}/2)^{2} + O(\tilde{k}_{z}^{4}).
\label{expandB4}
\end{eqnarray}
If $m=n$, then this becomes
\begin{equation}
B_{nn}^{(2)}(\tilde{k}_{z}) = {1\over \mu_{n}^{3/2}} - {3\over
4\mu_{n}^{5/2}}(\tilde{k}_{z}/2)^{2}
  + O(\tilde{k}_{z}^{4}).
\label{expandB5}
\end{equation}
Finally, for $B_{mn}^{(3)}$ we have
\begin{eqnarray}
B_{mn}^{(3)}(\tilde{k}_{z})& =&  - {3 h^{2}\over 4} \left({\tilde{k}_{z}\over
2}\right)
\int_{-1}^{1} dw \int_{0}^{1} dy\, {wy^{2}\over
  \left[ (\tilde{k}_{z}/2)^{2}(1-w^{2}y^{2}) + h(m-n)wy +
\mu_{(m+n)/2}\right]^{5/2}}
 \nonumber \\
 & = & \left[ {1\over (m-n)^{2}}\left({1\over\mu_{n}^{1/2}}
  - {1\over \mu_{m}^{1/2}} \right)\right.  \nonumber \\
   & & \qquad \left.  + {4\over h (m-n)^{3}}
   \left( \mu_{m}^{1/2} + \mu_{n}^{1/2} - 2\mu_{(m+n)/2}^{1/2}\right)\right]
   (\tilde{k}_{z}/2) + O(\tilde{k}_{z}^{3}).
\label{expandB6}
\end{eqnarray}
Note that this last integral is odd in $(m,n)$, in contrast to the first
two integrals, and vanishes when $m=n$.

\begin{center}
{FIGURE CAPTIONS}
\end{center}

{\sc FIG. 1}.  Illustration of a two-dimensional vortex lattice in
a nonuniform current.  The magnetic field is parallel to the $z$-axis,
normal to the film.  The three bold lines running parallel to the
$y$-axis are current contacts.  Equal currents are injected uniformly
along the two outer contacts and the total current is withdrawn
along the central contact.  This produces a current density, ${\bf J}$,
that is uniform along $y$, but nonuniform along $x$, as illustrated.
The forces on the vortex lattice due to the nonuniform current
elastically strain the lattice, producing the shear displacements,
${\bf u}$, shown.  These displacements are also uniform along $y$
but vary along $x$.  Outside of the outer contacts, ${\bf J=0}$
and ${\bf u}$ is uniform.

{\sc FIG. 2}. Zero field scaling functions at $d=3$.

{\sc FIG. 3}. Zero field scaling functions at $d=2$.

\clearpage

\end{document}